\title{The \textsc{Odd Even Delta} problem is \#P-hard}
\author{Giorgio Camerani}
\date{\today}

\documentclass[a4paper,11pt]{article}

\usepackage[noline,noend,ruled,linesnumbered]{algorithm2e}
\usepackage{amsmath}
\usepackage{amssymb}
\usepackage{anysize}
\usepackage[format=hang]{caption}
\usepackage{graphicx}
\usepackage{longtable}
\usepackage{lscape}

\marginsize{2cm}{2cm}{0.8cm}{2cm}

\begin{document}

\maketitle

\begin{abstract}
\hspace{-1.5em}Let $G=(V,E)$ be a graph. Let $k \leq |V|$ be an integer. Let $O_k$ be the number of edge induced subgraphs of $G$ having $k$ vertices and an odd number of edges. Let $E_k$ be the number of edge induced subgraphs of $G$ having $k$ vertices and an even number of edges. Let $\Delta_k = O_k - E_k$. The \textsc{Odd Even Delta} problem consists in computing $\Delta_k$, given $G$ and $k$. We show that such problem is \#P-hard, even on 3-regular bipartite planar graphs.   
\end{abstract}

\section{Introduction}
In this paper, we introduce the \textsc{Odd Even Delta} problem and prove it is \#P-hard, even on 3-regular bipartite planar graphs. The rest of the document is organized as follows:
\\\\
\begin{tabular}{p{2cm}p{14cm}}
\textbf{Section 2} & Defines the \textsc{Odd Even Delta} problem. \\
 & \\
\textbf{Section 3} & Shows the \#P-hardness of the \textsc{Odd Even Delta} problem on general graphs, by furnishing a polynomial-time Turing reduction from \textsc{\#Vertex Cover}. \\
 & \\
\textbf{Section 4} & Shows the \#P-hardness of the \textsc{Odd Even Delta} problem on 3-regular bipartite planar graphs, by invoking a known theorem. \\
 & \\
\textbf{Section 5} & Points to a working implementation, which is meant as a simple supporting tool to play with in order to practically verify the claim of this paper. \\
 & \\
\textbf{Section 6} & Ends the document by identifying what seems to be a promising direction of future work. \\
\end{tabular}

\section{Definition of the problem}
Let $G=(V,E)$ be a graph, and let $k \leq |V|$ be an integer. Let $O_k$ be the number of edge induced subgraphs of $G$ having $k$ vertices and an odd number of edges. Let $E_k$ be the number of edge induced subgraphs of $G$ having $k$ vertices and an even number of edges. Let $\Delta_k = O_k - E_k$. Our problem is defined as follows:
\\\\
\begin{tabular}{p{2cm}p{14cm}}	
\multicolumn{2}{l}{\Large{\textsc{Odd Even Delta}}} \\
 & \\
\large{\textsc{Input}} & Graph $G = (V,E)$, integer $k \leq |V|$. \\
 & \\
\large{\textsc{Output}} & $\Delta_k$ \\
\end{tabular}

\section{Proof of \#P-hardness on general graphs}
To prove that \textsc{Odd Even Delta} is \#P-hard on general graphs, we show a polynomial-time Turing reduction from \#\textsc{Vertex Cover}. Let $G=(V,E)$ be a graph. Let $\mathcal{C} = \{S \subseteq V : \forall \{u,v\} \in E\ u \in S \lor v \in S\}$ be the set of vertex covers of $G$. Let $I = \{ v \in V:\forall e \in E\ v \not\in e\}$ be the set of isolated vertices of $G$. Let $H = ( V - I, E )$ be the graph obtained from $G$ by removing its isolated vertices. Let $O_k$ be the number of edge induced subgraphs of $H$ having $k$ vertices and an odd number of edges. Let $E_k$ be the number of edge induced subgraphs of $H$ having $k$ vertices and an even number of edges. Let $\Delta_k = O_k - E_k$. Our reduction is expressed by the following equation:
\begin{equation}
\label{ReductionIsolated}
|\mathcal{C}| = 2^{|I|} \cdot (\ 2^{|V-I|} - \sum_{k=2}^{|V-I|} \Delta_k \cdot 2^{|V-I| - k}\ )
\end{equation}
When $G$ has no isolated vertices, $G = H$ and equation \ref{ReductionIsolated} simplifies to:
\begin{equation}
\label{ReductionNoIsolated}
|\mathcal{C}| = 2^{|V|} - \sum_{k=2}^{|V|} \Delta_k \cdot 2^{|V| - k}
\end{equation}
The reduction is polynomial-time, as \textsc{Odd Even Delta} is solved $|V| - 1$ times. Without loss of generality, we can assume that $G$ does not contain any isolated vertex. The remainder of this section is dedicated to explain why equation \ref{ReductionNoIsolated} holds.
\\\\
Let $\mathcal{U} = 2^V - \mathcal{C} = \{ S \subset V : \exists \{u,v\} \in E\ u \not\in S \land v \not\in S \}$ be the set of those subsets of $V$ that are not vertex covers of $G$. Clearly, the following holds:
\begin{equation}
\label{UCRelation}
|\mathcal{C}| = 2^{|V|} - |\mathcal{U}|
\end{equation}
For each edge $e = \{u,v\} \in E$, we define the set $\mathcal{U}_e = \{S \subset V : u \not\in S \land v \not\in S\}$. Observe how the existence of an edge $e \in E$ has the effect of putting outside $\mathcal{C}$, and thus in $\mathcal{U}$, all those $S \in \mathcal{U}_e$. The set $\mathcal{U}$ can be then expressed as:
\begin{equation*}
\label{UDefinition}
\mathcal{U} = \bigcup_{i = 1}^{|E|} \mathcal{U}_{e_i}
\end{equation*}
It is easy to see that for any two edges $e_1, e_2 \in E$, it is the case that $\mathcal{U}_{e_1} \cap \mathcal{U}_{e_2} \neq \varnothing$. As we have to determine the cardinality of a set defined as the union of non-disjoint sets, we invoke the inclusion-exclusion principle:
\begin{equation}
\label{InclusionExclusion}
|\mathcal{U}| = \biggl|\bigcup_{i = 1}^{|E|} \mathcal{U}_{e_i}\biggr| = \sum_{i = 1}^{|E|} (-1)^{i+1} \cdot \left( \sum_{1 \leq j_1 < \cdots < j_i \leq |E|} | \mathcal{U}_{e_{j_1}} \cap \cdots \cap \mathcal{U}_{e_{j_i}} | \right)
\end{equation}
Observe how the inner summation scans the edge induced subgraphs of $G$ having $i$ edges. While the outer summation scans the possible numbers of edges a subgraph may have, from just $1$ to $|E|$: note how those subgraphs having an odd number of edges give an additive contribution, whereas those having an even number of edges give a subtractive contribution.
\\\\
Let us focus on the generic term of the inner summation: the cardinality of the set $\mathcal{U}_{e_{j_1}} \cap \cdots \cap \mathcal{U}_{e_{j_i}}$. By applying the definition of $\mathcal{U}_e$, we can write:
\begin{equation*}
\mathcal{U}_{e_{j_1}} \cap \cdots \cap \mathcal{U}_{e_{j_i}} = \{ S \subset V : u_{j_1} \not\in S \land v_{j_1} \not\in S \land \cdots \land  u_{j_i} \not\in S \land v_{j_i} \not\in S \}
\end{equation*}
where $\{u_{j_i}, v_{j_i}\} = e_{j_i}$. It is now easy to see that the set $\mathcal{U}_{e_{j_1}} \cap \cdots \cap \mathcal{U}_{e_{j_i}}$ contains all and only those subsets of $V$ that do not contain any of the vertices of the subgraph induced by picking edges $e_{j_1} \cdots e_{j_i}$. 
\newpage
\hspace{-1.5em}Therefore its cardinality is expressed by the following equation:
\begin{equation*}
|\mathcal{U}_{e_{j_1}} \cap \cdots \cap \mathcal{U}_{e_{j_i}}| = 2^{|V|-k}
\end{equation*}
where $k$ is the number of vertices of the subgraph induced by picking edges $e_{j_1} \cdots e_{j_i}$. Hence the contribution given by the generic term of the inner summation does not depend on which nor on how many edges are picked, but only on the number of vertices of the subgraph induced by them. How many edges are picked is only relevant in the outer summation, to determine if such contribution has to be added or subtracted.
\\\\
Subgraphs with the same number of vertices give exactly the same contribution, regardless of their number of edges which affects only the sign of such contribution. By grouping subgraphs according to their number of vertices instead of according to their number of edges, we are thus able to rearrange equation \ref{InclusionExclusion} as follows:
\begin{equation}
\label{Rearrangement}
|\mathcal{U}| = \sum_{k = 2}^{|V|} O_k \cdot 2^{|V|-k} - \sum_{k = 2}^{|V|} E_k \cdot 2^{|V|-k} = \sum_{k = 2}^{|V|} \Delta_k \cdot 2^{|V|-k}
\end{equation}
The proof ends by observing that substituting equation \ref{Rearrangement} into equation \ref{UCRelation} leads to equation \ref{ReductionNoIsolated}.
\section{Proof of \#P-hardness on 3-regular bipartite planar graphs}
In order to prove that \textsc{Odd Even Delta} remains \#P-hard even on 3-regular bipartite planar graphs, it is sufficient to invoke the following theorem \cite{XiaZhao}:
\begin{quotation}
\emph{\textsc{\#3-Regular Bipartite Planar Vertex Cover} is \#P-complete}
\end{quotation}
\section{Implementation}
A Java implementation is available at \texttt{gcamerani.altervista.org}. It is composed by two blocks: the first block computes $\Delta_k$ for each $k \in [2, |V|]$, while the second block computes $|\mathcal{C}|$ according to equation \ref{ReductionNoIsolated}. As this implementation is only meant as a didactical support to practically verify equation \ref{ReductionNoIsolated}, the first block has been left as unefficient as possible: it merely enumerates all the edge induced subgraphs of $G$.

\section{Future work}
The \#P-hardness of \textsc{Odd Even Delta} on 3-regular bipartite planar graphs seems to be an interesting research avenue, deserving further exploration. More precisely, such exploration can be conducted in two opposite ways. Those who are trying to prove \#P = FP can try to focus their efforts in devising an efficient algorithm to compute $\Delta_k$, possibly by leveraging the bipartiteness restriction: maybe there is a clever manner to quickly obtain $\Delta_k$ without computing $O_k$ and $E_k$ at all, or maybe $O_k$ and $E_k$ are themselves efficiently computable. On the other hand, those who are trying to prove \#P $\neq$ FP can try to channel their investigations on the intrinsic reason for which even computing just $\Delta_k$ may forcedly require us to consider a nevertheless large fraction of the exponentially many edge induced subgraphs of $G$.

\end{document}